\documentclass[conference]{IEEEtran}
\IEEEoverridecommandlockouts
\usepackage{cite}
\usepackage{amsmath,amssymb,amsfonts}
\usepackage{algorithmic}
\usepackage{graphicx}
\usepackage{textcomp}
\usepackage{xcolor}
\usepackage{array}
\def\BibTeX{{\rm B\kern-.05em{\sc i\kern-.025em b}\kern-.08em
    T\kern-.1667em\lower.7ex\hbox{E}\kern-.125emX}}

\begin{document}

\title{Forensic Analysis of Third Party Location Applications in Android and iOS\\
}

\author{\IEEEauthorblockN{Jason Bays}
\IEEEauthorblockA{Department of Computer and Information Technology\\
Purdue University\\
West Lafayette, Indiana \\
bays0@purdue.edu}
\and
\IEEEauthorblockN{Umit Karabiyik}
\IEEEauthorblockA{Department of Computer and Information Technology\\
Purdue University\\
West Lafayette, Indiana \\
umit@purdue.edu}}

\maketitle

\begin{abstract}
Location sharing applications are becoming increasingly common. These applications allow users to share their own locations and view contacts' current locations on a map. Location applications are commonly used by friends and family members to view Global Positioning System (GPS) location of an individual, but valuable forensic evidence may exist in this data when stored locally on smartphones. This paper aims to discover forensic artifacts from two popular third-party location sharing applications on iOS and Android devices. Industry standard mobile forensic suites are utilized to discover if any locally stored data could be used to assist investigations reliant on knowing the past location of a suspect. Security issues raised regarding the artifacts found during our analysis is also discussed.
\end{abstract}


\section{Introduction}
Mobile applications allowing the sharing and retrieval of GPS location are now very common on the iOS App Store and Google's Play Store. These allow users of smartphones to grant access to their location to others, providing for on-demand GPS access. This is often sold as a safety application, such as a parent ensuring their child makes it to school. Another use is to allow convenience among friend groups when meeting up at a location. 

Because of the always-on nature of these applications, valuable forensic data may exist on the local storage of devices with them installed. This could be of use to law enforcement. Some investigations rely on location-based evidence to corroborate a story or track a suspect's movements. There may be cases where the suspect's phone is not available, but one of their accomplice's devices is available. Even if the user logs out of the application, location data may still be recoverable. Previous forensics research has looked at general GPS artifact locations on iOS and Android operating systems, as well as native applications' location tracking history \cite{rose2012ubiquitous}. However, there is a lack of research on third-party location sharing applications utilized by many families and friend groups.

The target devices for this study are one device using iOS and one device using Android operating system (OS). These OSs were chosen because of their large market share; according to the International Data Corporation, Android has 85.0\% of the global market, with iOS in second at 14.7\% \cite{idc}. These two operating systems overwhelmingly represent the smartphone market. They are also well-supported by forensic tools and processes. Authors in \cite{levinson2011third} have documented methods for third-party smartphone forensic acquisition, and authors in \cite{ntantogian2014evaluating} presented evaluation of privacy when forensically analyzing an Android device.

The two applications chosen for this study are Life360 and iSharing. These applications were chosen because of their popularity on the market and their support for both operating systems. As of January 2019, Life360 was the seventh most popular application in the Social Networking category for iOS, with iSharing as a top choice for GPS applications on Android. Both applications support GPS tracking of friends and family, and they both advertise historical viewing of location data. The operation of location services on mobile devices is well-understood by previous studies \cite{kushwaha2011location,strawn2009expanding}. Based on this research, it is reasonable to assume GPS data could be found by a forensic analysis tool.

In this paper, we demonstrated our forensic analysis on  analyzing aforementioned applications installed in both Android and iOS devices. The acquisition process is explained in detail with challenges faced. We presented the absence or presence of GPS data as well as lists of users on the applications, and GPS tiles. Moreover, the files acquired from the devices are examined for completeness compared to the populated data. The methods presented in this paper will assist law enforcement in handling location-sensitive investigations when the location sharing applications are found during the investigation.

\section{Review of Relevant Literature}\label{sec:literature}
Both iOS and Android devices utilize the  GPS in order to receive and transmit location data \cite{strawn2009expanding}. GPS was first developed by the United States Department of Defense in 1978. There are 24 main satellites orbiting the Earth which send radio signals to the ground. GPS-enabled devices read data in order to provide accurate location data \cite{strawn2009expanding}. By utilizing GPS data, smartphones are able to provide services such as turn-by-turn navigation and location tracking applications such as the ones described in this study. Apple includes a first-party location tracking application in iOS. Known as Find My Friends, this application allows users to share and receive locations of other users, similar to the applications analyzed in this present study. Third-party applications are downloaded from the App Store on iOS and the Google Play store on Android. At the time of this article's writing, two of the most popular applications for location sharing are iSharing and Life360. These applications were chosen for their popularity, their advertised features of location history, and their availability on both platforms. Life360 currently has over 222,000 ratings on the App Store, and it was featured as App of the Day on the store, being sold as an application which promotes safety. Analyzing these applications appears representative of the current most popular location tracking applications available on the markets. 

\subsection{iOS}
Previous research has shown the existence of GPS evidence on iOS. In 2012, iOS received media attention for the existence of a file which permanently stored a user's location history \cite{rose2012ubiquitous}, which could be combined with mapping software to display everywhere a user had been since purchasing their phone. Apple later patched this behavior, claiming the file was behaving abnormally and the intended behavior had been to only store data for two weeks.  

Location data on iOS is processed and received using the \textit{CoreLocation} framework. CoreLocation is a built in component of the iOS Application Programming Interface (API), available for developers to use when producing iOS applications. Many popular applications and services, such as Google Maps, use this framework. iOS is able to use the ``latitude, longitude, and altitude of the user's device, along with the level of accuracy to which this information is known, and the direction in which the device is moving \cite{allan2012geolocation}." CoreLocation tracks this information using either the cellular network, the WiFi network location, or GPS, and each one of these methods is more accurate than the previous. Therefore, regardless of positioning method, the location will be stored in one central area. Both GPS and cellular data positioning use triangulation. Triangulation connects to at least three GPS satellites or cell towers and using the signal strength and positioning of each, uses this to determine device location.

Another possible data source on iOS devices is the App Caches folder \cite{reiber2016mobile}. Each application has its own set of cache folders where it can store temporary data. This data can vary greatly among each application, but it can contain valuable forensic data. This is not typically found in a standard iTunes backup. ``Because the Caches folder is not contained within an iOS standard iTunes backup, the mobile forensic solution must use the house\_arrest and file\_relay services for recovery and collection \cite{reiber2016mobile}." It is possible the Life360 and iSharing cache folders contain cached data of users' locations. Even if coordinates are not found, discovery of map tiles could reveal where other users have been in a more generalized area. 

\subsection{Android}

Research has also been conducted on location forensics on Android. Location based services provide for location tracking on Android, which keep records on the user's current and previous locations \cite{kushwaha2011location}. Location based services on Android also provide for notifications when users move in and out of an area. GPS functionality is simple to enable in Android, often requiring only a single XML file to configure. The Android and iOS variants of iSharing and Life360 should function identically, as GPS functionality is now common and standard on both operating systems. Android is also vulnerable to leaking GPS information without the user's knowledge. Common enterprise applications, including one built into devices manufactured by HTC, were found to store unencrypted GPS location \cite{wei2012malicious}. These unencrypted files should easily be found by modern forensic tools. Previous research has also shown possible forensic artifact locations for Android location data \cite{spreitzenbarth2012comparing}. Android stores the latest cell towers and WiFi routers in the system folder's cache files. The web browser stores GPS locations in \textit{CachedGeopositions.db}. Third party applications were also found to store GPS locations in their database files. Twitter, for example, stored latitude and longitude of status messages in a file labeled author\_id.db. These coordinates were stored despite Twitter not being known as a location sharing application. Android applications also commonly contain cache folders, which should also be examined. Map tiles, GPS coordinates, and lists of affiliated users are all potential artifacts which could be found in the Android application's folder.  

\subsection{Acquisition Methods}

Multiple types of extractions can be used to produce a forensic image of smartphones. A forensic image is a copy of the data contained on the device. A forensic investigator should aim to obtain the most complete record of data possible while avoiding altering evidence. Cellebrite, a commonly used mobile forensics tool vendor, described the two main types of extraction - physical and logical \cite{explaininghappens}. A third form, file system extraction, is a more thorough variant of a logical extraction. The most complete type of acquisition is a physical acquisition, which takes a bit-for-bit copy. It captures the entirety of the device's storage, including unallocated and slack space. This provides for the possibility of recovering deleted and hidden files. Physical acquisition, however, often requires invasive and destructive methods such as JTAG to obtain \cite{norouzizadeh2016investigating}. Industry tools such as Cellebrite UFED Analyzer and Magnet AXIOM can perform physical acquisitions on some older iOS and Android devices and OS versions, but it is not typically possible on newer versions of each operating system. Logical extractions use the operating system's built-in API to obtain data \cite{explaininghappens}. While commonly obtainable, logical extractions are limited to what the manufacturer makes available. However, they can still contain a large amount of valuable data, and sometimes are the only option for an acquisition. A file system acquisition, commonly available for rooted Android phones or jailbroken iOS devices, obtain the entire filesystem, including internal system folders. This can result in additional data being found. However, jailbreaking an iOS or rooting an Android device is a process which changes the file system and potentially alters the evidence. It is important to document why this is necessary when performing an investigation. Rooting an Android device consists of gaining full unrestricted access to the file system of the device \cite{sun2015android}.

An additional option for an acquisition on iOS is analyzing an iTunes backup file \cite{husain2010simple}. Most mobile forensic tools can open and analyze backup files, which has been found to contain some location-related files. For example, the ``Map" directory on an iOS device contains a \textit{history.plist} file, which lists GPS locations. While this study will have access to a full device image, it is important to note investigators may rely on the backup file if the device itself is unrecoverable. 

Overall, the most complete method for obtaining an acquisition on an iOS or Android device will be a physical acquisition if possible, and a logical acquisition if not. The iOS device will need to be jailbroken in order to achieve this on modern iOS versions, and Android will require root to be obtained. These methods give access to lower level system functions \cite{barmpatsalou2013critical}, allowing more complete acquisitions. The latest jailbreak available for iOS, known as Electra, supports iOS versions up to 11.3.1 \cite{afonin_2018}. This iOS version was released on April 24, 2018. No jailbreak currently exists for any variant of iOS 12. Methods of obtaining root vary greatly by device for Android. It will be most valuable to use the latest version of the operating systems available for each of these methods.  

In this section, we provided an overview of the literature relevant to mobile phone acquisition and GPS data. Previous research has been performed on GPS applications, and documented methods exist for obtaining a forensic image on both iOS and Android. Multiple acquisitions, both physical and logical, should be performed. The cache folders should be examined for forensic evidence. The next section will outline the specific methodology to be performed for both iOS and Android.

\section{Methodology}

This study will follow a standard forensic model to acquire and analyze images from the devices. Three devices were utilized and chosen for this study - an iPhone 7 running iOS 12.0.1, an iPhone 5 running iOS 10.3.3, and a Samsung Galaxy S7 running Android 8.0. The iPhone 7 and Galaxy S7 were chosen to ensure the most recent software was analyzed for the study. The iPhone 5 was utilized to ensure compatibility with a jailbreak, as jailbreaking often discovers files and caches not acquirable by other methods. 

The tools used for acquisition and analysis in this study include Cellebrite UFED 4PC 7.5.0.845 and Magnet AXIOM 2.6.0.11689. These are both standard forensic tools used by large law enforcement agencies worldwide. Commercial mobile forensic tools are often compared for effectiveness by research, such as \cite{alhassan2018comparative}. These tools are both capable of performing all acquisition types of supported mobile devices. 

The forensic workstation utilized for this study was a Dell Optiplex 7060 running Windows 10 Education 64-bit Build 17134. The workstation had 16GB of RAM and utilized an Intel Core i7-8700 CPU.

Two new iCloud accounts were created for this study, as well as a new Gmail account. The iCloud accounts were used to setup the iPhone devices as a new device. The Gmail account was used to setup the Galaxy S7 device as a new device. The devices were configured to factory default settings. The only addition was configuring the WiFi networks used to track location. The outline for the methodology is as follows:

\begin{itemize}
\item Wipe and restore all devices to factory defaults.
\item Configure devices using empty iCloud and Gmail accounts.
\item Install latest version of iSharing and Life360.
\item Perform a baseline logical acquisition of each device.
\item Populate the devices with data.
\item Allow each device to track locations for 72 hours.
\item Perform a second logical acquisition.
\item Jailbreak the iPhone 5.
\item Root the Samsung Galaxy S7.
\item Perform a full file system acquisition of the Galaxy S7 and iPhone 5.
\item Analyze results in UFED Physical Analyzer and AXIOM Examine.
\end{itemize}

\subsection{Data Population}

In accordance with National Institute of Standards and Technology (NIST) standards \cite{ayers2018quick}, each phone was populated with sample data. The iSharing and Life360 applications were installed on a personal device which was in frequent location movement, an iPhone XS Max. Both applications were also installed on each research device. Profiles were populated with name, profile picture, and home locations. Each application was populated with all other device as location sharing contacts.  

The devices were allowed to connect and track location data for 72 hours. During this period of time, the chat and voice call functions of each application were utilized. This simulates how a typical user may use the application, and ensures a sufficient amount of potential data exists. During this period of time, GPS and WiFi functionality was enabled.

\subsection{Analysis}

Logical acquisitions were performed for the baseline and first analysis acquisition. A logical acquisition was chosen because it does not modify significant amount of data on the device, and it is widely supported across tools and devices. The application folders were identified on each device, as these are of particular interest. On Android, the file path for the iSharing application was found to be \textit{/apps/com.isharing.isharing/db/crash\_reports}. The file path for Life360 was \textit{/sdcard/Android/data/com.life360.android.safetymapd}. On iOS, the file path for iSharing was \textit{/Applications/com.isharing.iSharing-Lite}. The file path for Life360 was Location's \textit{/Applications/com.life360.safetymap}.

These acquisitions were taken using both Cellebrite UFED4PC and Magnet AXIOM. The backup method was selected for Android. Once these acquisitions had been performed and archived, the iPhone 5 was jailbroken using the h3lix Jailbreak. This allows full filesystem access to the device. The Galaxy S7 was rooted using Samsung's ODIN tool. ODIN is an internal application used to flash new firmware onto a device \cite{alendal2018forensics}. Once the custom firmware was loaded, root access was confirmed by browsing the phone's filesystem.

Once these steps were complete, physical acquisitions were attempted for each device. The iPhone failed physical acquisition in both tools. This is because the h3lix jailbreak, the tool available for devices running iOS 10, does not include OpenSSH by default. OpenSSH is required in order to obtain an acquisition in Magnet AXIOM. Cellebrite's UFED4PC also refused to obtain the acquisition. However, a manual acquisition is possible, by simply manually browsing through the phone's filesystem. Important files are outlined in Section \ref{Results}, which can be found by manually acquiring the phone. 

The Samsung Galaxy S7 was able to be acquired in Magnet AXIOM. Magnet AXIOM recognizes the rooted status of the device, and uses this to gain privileged access. Upon completion, the full file system was decoded and available for browsing. Cellebrite UFED4PC recognized the device as rooted, but did not allow a physical acquisition. An Android Debugging Bridge (ADB) physical acquisition was attempted, but Cellebrite did not support the security level of the operating system. Other devices, or other root methods, may be able to be acquired successfully in Cellebrite UFED4PC.

\section{Results}\label{Results}

Upon analysis of the forensic images, artifacts of forensic interest were found on all devices. A summary of whether an artifact is available or not is depicted in Table \ref{Summary}. iOS provided access to the contact list in both applications, as well as a large amount of other location data from Life360. Android presented results largely the opposite, with most location data recovered from the iSharing application. Upon jailbreaking the iOS 10 device, a large amount of additional information, including the user's password in plaintext, was found in the iSharing application folder. Upon rooting the Android device, more location and messaging data was found from Life360's data folder.

\begin{table}[]
\centering
\scriptsize
\caption{Summary of Artifact Availability}\label{Summary}
\setlength{\extrarowheight}{3pt}
\begin{tabular}{l|c|c|c|c|}
\cline{2-5}
 & \multicolumn{2}{c|}{\textbf{Android}} & \multicolumn{2}{c|}{\textbf{iOS}} \\ \cline{2-5} 
 & \textbf{iSharing} & \textbf{Life360} & \textbf{iSharing} & \textbf{Life360} \\ \hline
\multicolumn{1}{|l|}{\textbf{Contact Lists}} & Yes & No & Yes & Yes \\ \hline
\multicolumn{1}{|l|}{\textbf{User Information}} & Yes & No & Yes & No \\ \hline
\multicolumn{1}{|l|}{\textbf{User Password}} & No & No & Yes & No \\ \hline
\multicolumn{1}{|l|}{\textbf{User GPS Coordinates}} & Yes & No & No & Yes \\ \hline
\multicolumn{1}{|l|}{\textbf{Contact GPS Coordinates}} & Yes & No & Yes & No \\ \hline
\multicolumn{1}{|l|}{\textbf{Messages}} & No & Yes & No & No \\ \hline
\end{tabular}
\end{table}

\subsection{iOS}

\begin{table*}[]
\centering
\caption{iOS 12 Artifact Locations}\label{iOS12}
\begin{tabular}{|l|l|l|}
\hline
\textbf{Artifact Type}           & \textbf{Application} & \textbf{Location}         \\
\hline
Contact List            & iSharing    & iPhone/Applications/com.isharing.iSharing-Lite/Documents/isharing.db                                 \\
Contact List            & Life360     & iPhone/Applications/com.life360.safetymap/Library/Application Support/SafetyMap/L360Model.sqlite     \\
Contact GPS Coordinates & iSharing    & iPhone/Applications/com.isharing.iSharing-Lite/Documents/isharing.db                                 \\
User GPS Coordinates    & Life360     & iPhone/Applications/com.life360.safetymap/Documents/LogFile.txt                                    \\
Remembered Locations    & Life360     & iPhone/var/root/Library/Caches/locationd/consolidated.db                                            \\
Remembered Locations    & Life360     & iPhone/Applications/com.life360.safetymap/Library/Application Support/SafetyMap/L360Model.sqlite    \\
\hline
\end{tabular}
\end{table*}

Contact lists were found for both applications in iOS. These contact lists can include the user's name, email address, phone number, and last seen coordinates. On Life360, this is found in the \textit{L360Model.sqlite} database within the \textit{ZL360Member} table. It is important to note this database includes the contacts' full names, email addresses, and phone number, as shown in \ref{GPS_FIGURE_2}. On the iSharing application, this is found in the \textit{iSharing.db} database within the  Person table. This table also lists accuracy of the GPS measure as well as the last battery level recorded for each user. 

As mentioned in Section~\ref{sec:literature}, iOS uses the CoreLocation framework to consolidate GPS information. CoreLocation's \textit{consolidated.db} file was analyzed, and it was found to contain marked locations for Life360. This could include locations the user has deemed important, and the user has set the application to provide notifications when other users enter or exit the area.

\begin{table*}[]
\centering
\caption{iOS 10 (Jailbroken) Additional Artifact Locations (all locations are under directory /private/var/mobile)}\label{iOS10_Jail}
\footnotesize
\begin{tabular}{|l|l|l|}
\hline
\textbf{Artifact Type}     & \textbf{Application} & \textbf{Location}  \\
\hline
User Password     & iSharing    & /Containers/Data/Application/(application ID)/Library/Preferences/com.iSharing.iSharing-Lite.plist  \\
User Phone Number & iSharing    & /Containers/Data/Application/(applicaiton ID)/Library/Preferences/com.iSharing.iSharing-Lite.plist  \\
User Home Address & iSharing    & /Containers/Data/Application/(application ID)/Library/Preferences/com.iSharing.iSharing-Lite.plist \\ 
\hline
\end{tabular}
\end{table*}

When a jailbroken iOS device was analyzed, additional artifacts of significant forensic interest were found for the iSharing application. The \textit{iSharing-Lite.plist} file contains many entries setting the configuration of the application. Of note, the user's username and password are stored in plaintext in this file. In addition, the user's defined home address is stored in this file as well, as well as the registered phone number. This file contains nearly every piece of the user's profile.

\subsection{Android}

\begin{table*}[]
\centering
\caption{Android (Unrooted) Artifact Locations}\label{Android}
\scriptsize
\begin{tabular}{|l|l|l|}
\hline
\textbf{Artifact Type}           & \textbf{Application} & \textbf{Location}                                                                                                 \\
\hline
User Email              & iSharing    & Samsung GSM\_SM-G930U Galaxy S7.zip/apps/com.isharing.isharing/sp/ISHARING\_PREFS.xml                    \\
Contact List            & iSharing    & Samsung GSM\_SM-G930U Galaxy S7.zip/apps/com.isharing.isharing/db/isharing\_talk                         \\
Contact GPS Coordinates & iSharing    & Samsung GSM\_SM-G930U Galaxy S7.zip/apps/com.isharing.isharing/db/isharing\_talk                          \\
User GPS Coordinates    & iSharing    & Samsung GSM\_SM-G930U Galaxy S7.zip/apps/com.isharing.isharing/db/com.hypertrack.common.device\_logs.db  \\
Profile Pictures        & Life360     & Samsung GSM\_SM-G930U Galaxy S7.zip/sdcard/Android/data/com.life360.android.safetymapd/cache            \\
\hline
\end{tabular}
\end{table*}

On Android, Life360 was well-protected on an unrooted device, revealing only the user's profile picture. This was stored as a JPG file in the application directory. No user profile information, nor any information about associated contacts, was found for Life360 on Android. However, iSharing contained many of the same artifacts found on the iOS version. The \textit{isharing\_talk} database contained many elements of the user profiles, including name and GPS coordinates. The user email was found in the \textit{ISHARING\_PREFS.xml} file, which contains the settings and configuration for the application. 

\begin{table*}[]
\centering
\caption{Android (Rooted) Additional Artifact Locations}\label{Android_Root}
\scriptsize
\begin{tabular}{|l|l|l|}
\hline
\textbf{Artifact Type }                      & \textbf{Application} & \textbf{Location}                                                                                        \\
\hline
Messages and Associated Coordinates & Life360     & Samsung GSM\_SM-G930U Galaxy S7.zip/data/com.life360.android.safetymapd/databases/messaging.db \\
\hline
\end{tabular}
\end{table*}

\begin{figure*}[!t]\label{GPS_FIGURE}
\centering
\includegraphics[width=0.7\textwidth]{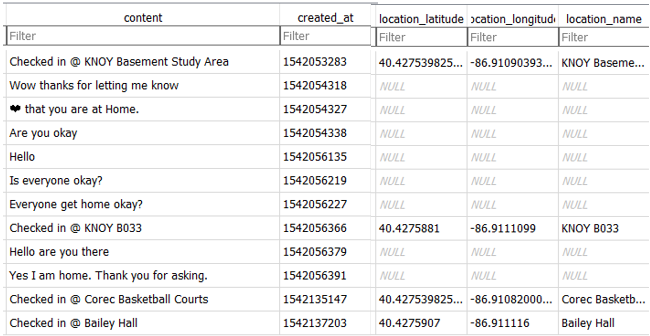}
\caption{Plaintext messages stored in messaging.db}
\label{fig:messages}
\end{figure*}

\begin{figure}[!t]\label{GPS_FIGURE_2}
\centering
\includegraphics[width=0.49\textwidth]{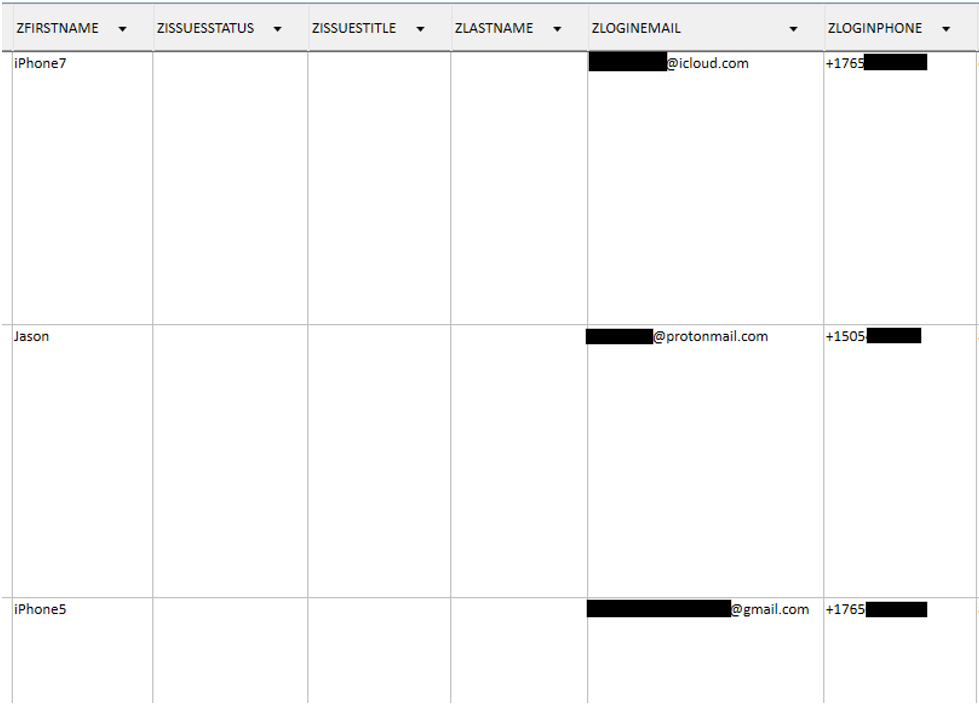}
\caption{User information stored in plaintext in Life360's database}
\label{fig:messages2}
\end{figure}

Upon rooting the Android device, more artifacts were found for Life360. The \textit{messaging.db} database contained many user interaction components. Life360 allows for messaging of other contacts, and all messages were found in plaintext for this application. The GPS location of where messages were sent was found for some messages, particularly the automated messages Life360 sends when someone arrives or leaves home. For an example of this type of information in the database, see Fig.~\ref{fig:messages} and Fig. ~\ref{fig:messages2}. 

\section{Discussion and Conclusion}

Based on the results presented, it was found that artifacts of forensic importance could be located in both applications. Digital forensics investigators may be able to recover GPS information of both the user and the user's contacts. The amount of information available will depend on the operating system and whether the full file system is available.

The first step an investigator should take when encountering one of these applications is to determine the operating system of the device. The investigator should then choose the most appropriate acquisition method for the device. The specific forensic tool should not have a large impact, as the results found were the files existent on the device's normal filesystem. However, it is recommended to use a tool which can decode and visually display databases, as many of the relevant artifacts exist in a database file. If it is an iOS device, the investigator should check for whether the device is jailbroken. If the device is not jailbroken, the results in Table \ref{iOS12} are the most appropriate places to look for evidence. If the device is jailbroken, one should also check the locations in Table \ref{iOS10_Jail}. If the device is an Android device, the results in Table \ref{Android} should be examined. If the device has been rooted, and the user is using Life360, the results in Table \ref{Android_Root} should be examined next. 

A digital investigator should only access the level of data they require before rooting or jailbreaking a phone. While jailbreaking will result in additional artifacts, jailbreaking also changes data permanently on an iOS device \cite{chang2015jailbroken}. The same is true for rooting on Android. While reversing a jailbreak or root may be possible, all artifacts from the logical acquisition should be considered first. 

Of particular note was the presence of the user's username and password in plaintext found on the jailbroken iOS device. These credentials would allow access to the entire application and all associated data. This would allow digital investigators to login to the application even if the image was a backup or the phone had been restored.

The amount of information found overall supports the research question. The data available could assist in solving investigations related to kidnapping, missing persons, or organized criminal rings. It is possible to see who a user is associated with as well as their last recorded location. Commonly accessed locations and reminder locations are stored as well. Finally, on Android, all messages sent between associated users are viewable in plaintext. These are all artifacts which can assist an investigator in a case where location data is vital.

This work also displays the importance of privacy and security awareness among users of the applications. Users may not realize their information is being stored locally on their contacts' phones, including their email address and last known location. This could pose a risk of their information being released if a phone is stolen or sold without first being erased. In addition, the user password for iSharing application is not securely stored since it can be recovered in plaintext.

Future work would include extending this work to other common location sharing applications and testing on additional devices. Additional analysis could be done in regards to the social media integration these applications provide in order to discover if any social media artifacts are stored. This study was limited to two iOS devices and one Android device; it is possible additional devices may further validate the results or find additional artifacts. Another possible research area would involve looking at the paid, premium variants of these applications. The premium versions advertise longer location history, which may result in additional artifacts. This research focused on the free, unpaid versions. 

In summary, GPS artifacts can be found on both iOS and Android devices when using the Life360 or iSharing applications. These artifacts range from contact lists to GPS coordinates and saved locations. On devices with access to the full file system, more information can be recovered, such as plaintext messages and possibly full account information. This data can assist a law enforcement agency during the course of a digital investigation if location of an individual or an individual's associates are important.

\bibliography{bibliography}
\bibliographystyle{ieeetr}

\end{document}